\documentstyle[twocolumn,eqsecnum,aps,epsfig]{revtex}
\begin{document}
\title{Directed and Elliptic Flow}
\author{Sven~Soff$^a$, Steffen~A.~Bass$^b$, Marcus~Bleicher$^a$, Horst~St\"ocker$^a$, Walter~Greiner$^a$}
\address{$^a$Institut f\"ur Theoretische Physik, J.W.\ Goethe Universit\"at,\\
Robert-Mayer Str.\ 10, 60054 Frankfurt, Germany}
\address{$^b$Department of Physics, Duke University\\
Durham, NC 27708, USA}
\date{\today}
\maketitle   
\begin{abstract}
We compare microscopic transport model calculations to 
recent data on the 
directed and elliptic flow of various hadrons 
in $2\-- 10\,A\,$GeV Au+Au and Pb$\,(\,158\,A\,{\rm GeV}\,)\,$Pb collisions. 
For the Au+Au excitation function a transition from the squeeze-out to 
an in-plane enhanced emission is consistently described 
with mean field potentials corresponding to one incompressibility. 
For the Pb$\,(\,158\,A\,{\rm GeV}\,)\,$Pb system the elliptic flow 
prefers in-plane emission both for protons and pions, 
the directed flow of protons is opposite to that of the pions, which 
exhibit anti-flow. 
Strong directed transverse flow is present for protons and $\Lambda$'s in 
Au($6\,A\,$GeV)Au collisions as well. 
Both for the SPS and the AGS energies the agreement between data and 
calculations is remarkable.

\end{abstract}
\narrowtext

Recently, it has been reported on an enormous amount of new detailed 
data on the collective flow in 
relativistic heavy ion collisions \cite{pinkenburg99,appelshaeuser98,aggarwal98}.

The excitation function of transverse collective flow is
the earliest predicted signature for probing
compressed nuclear matter \cite{scheid68a}.
Its sensitivity to the equation of state (EoS) can be used to search for abnormal matter states
and phase transitions \cite{hofmann76a,stoecker86a}.

In the fluid dynamical approach, the transverse collective flow is directly
linked to the pressure $P(\rho, S)$  (depending on the density $\rho$
and the entropy $S$) of the matter  in the reaction
zone:

One can get a physical feeling for
the generated collective transverse momentum $\vec{p}_x$ by writing
it as an integral of the pressure acting on a surface and over
time \cite{stoecker81a}:
\begin{equation}
\label{pxeqn}   
\vec{p}_x \,=\, \int_t \int_A P(\rho,S) \, {\rm d}A \, {\rm d}t\,.
\end{equation}
Here d$A$ represents the surface element between the participant and
spectator matters and the total pressure is the sum of the
potential pressure and the kinetic pressure:
The transverse collective flow depends directly on the equation of state,
$P(\rho,S)$.

Collective flow had originally been  predicted by nuclear shock wave models and 
ideal fluid dynamics (NFD)
\cite{scheid68a,stoecker80a,stoecker82a,buchwald84a}.
Microscopic models such as VUU
(Vlasov Uehling Uhlenbeck), and QMD (Quantum Molecular Dynamics) have
predicted smaller flow than ideal NFD. These microscopic models agree roughly 
with viscous NFD \cite{schmidt93a} and
with data \cite{moli84,moli85,gyulassy82,hartnack89a,ch92,peilert94}, which discovered 
flow first at the
BEVALAC \cite{gus84,do86,gut89} for charged particles by the Plastic-Ball
and Streamer Chamber  collaborations \cite{renfordt84a}, and at SATURNE by the
DIOGENE collaboration \cite{gosset90a,dem90}. 
It has been studied extensively at GSI by the FOPI
\cite{ramillien95a,herrmann96a}, LAND \cite{leifels93a}, TAPS
\cite{kugler94a}, and KaoS \cite{brill96a} collaborations, and by the 
EOS-TPC collaboration at LBNL \cite{chance96a} and at MSU 
\cite{westfall93}.

Two different signatures of collective flow have been predicted:
\begin{itemize}
\item[a)] The {\em bounce--off} \cite{stoecker80a} of compressed matter
{\em in the reaction plane} and
\item[b)] the
{\em squeeze--out} \cite{stoecker82a} of the participant matter
{\em out of the reaction plane}.
\end{itemize}
The most strongly stopped, compressed matter
around mid-rapidity is seen directly in the {\em squeeze--out} \cite{ch92}.
A strong dependence of these collective effects
on the nuclear equation of state
is predicted \cite{ch92}. For higher beam energies, however,
projectile
and target spectator decouple quickly from the reaction zone, giving
way to a preferential emission of matter in the reaction plane,
even at mid-rapidity \cite{ollitrault93a}.
An excitation function of the {\em squeeze--out} at midrapidity shows 
the transition from out of plane enhancement to preferential
in-plane emission. 


At 10.6 $A\,$GeV collective flow has recently been discovered
by the E877 collaboration \cite{barrette94c,barrette96a} 
by measuring  
d$v_1/{\rm d}\eta =
{\rm d}(\langle E_x\rangle/\langle E_T\rangle)/{\rm d}\eta$ for different
centrality bins.
The EOS group has
measured the flow excitation function for Au+Au  at the AGS in the
energy range between 2.0 and 8 GeV/nucleon \cite{liu98a}. 
Their data show a smooth decrease in $\langle p_x \rangle$ 
from 2 to 8 GeV/nucleon and are corroborated by measurements of the 
E917 collaboration at 8 and 10.6 GeV/nucleon \cite{ogilvie98a}. 

The EOS collaboration has also measured a {\em squeeze-out} excitation
function (sometimes also termed ``elliptic flow'' \cite{ollitrault93a}),
indicating a transition from out-of-plane to in-plane enhancement around
5 GeV/nucleon \cite{liu98a}.

At CERN/SPS, the first observations of the predicted directed transverse flow
component \cite{keitz91} have been reported by the WA98 collaboration
\cite{nishimura98a,aggarwal98}
using the Plastic Ball detector located at target rapidity for event plane
reconstruction. They show a strong directed flow signal for protons
and ``antiflow'' for pions, both enhanced for particles with high transverse
momenta. Similar findings have also been reported by the NA49 collaboration,
which due to their larger acceptance allows for a more detailed
investigation \cite{appelshaeuser98}.

Due to its direct dependence on the EoS, $P(\rho,T)$, flow excitation
functions can provide unique information about phase transitions:
The formation of abnormal nuclear matter, e.g., yields a reduction
of the collective flow \cite{hofmann76a,peilert94}.
A directed flow excitation function as   
signature of the phase transition into the QGP has been proposed
by several authors 
\cite{hofmann76a,stoecker86a,vanhove82a,amelin91,rischke95b,brachmann97,shuryak95}.

A microscopic analysis showed that the existence of
a first order phase transition
can show up as a reduction in the directed transverse
flow \cite{ch92,peilert94}.

For first order phase transitions, the pressure remains constant (for $T={\rm const}$)
in the region of the phase coexistence. This results in vanishing 
shock velocities $v_f=0,\, v_s=0$ and 
velocity of sound $c_s= \sqrt{\partial p/\partial \varepsilon}$ 
\cite{hofmann76a,stoecker86a}.   

The expansion of the system is driven by the pressure gradients, therefore
expansion depends crucially on $c_s^2$. Matter in the  mixed phase
expands less rapidly than a hadron gas or a QGP at the same energy density
and entropy.
In case of rapid changes in the EoS without phase transition, the
pressure gradients are finite, but still smaller than for an ideal gas
EoS, and therefore the system expands more slowly \cite{kapusta85,gersdorf87}.

This reduction of $c_s^2$ in the transition region is commonly referred to
as {\em softening} of the EoS. 
Here the
flow will temporarily slow down (or possibly even stall).
This hinders the deflection of spectator matter
(the {\em bounce--off}) and, therefore, causes  
a reduction of the directed transverse
flow \cite{hofmann76a,shuryak95,amelin91,rischke95b}
in semi-peripheral collisions.   
The softening of the EoS should be observable in the excitation function of the
transverse directed flow of baryons.

An observation of the predicted
local minimum in the excitation function of the directed transverse flow
\cite{hofmann76a,rischke95b,rischke96a}
would be an important discovery, and an unambiguous signal for a
phase transition in dense matter.
Its experimental measurement would serve as strong evidence for
a QGP, if that phase transition is of first order.

An illustration of the in-plane elliptic flow is given by the following picture: 
Two colliding nuclei 
create a stopped overlap region. At higher bombarding 
energies ($E_{\rm lab} \ge 10 \,A\,$GeV) the spectators 
leave rapidly this interaction zone. The remaining interaction zone 
expands almost freely, where the surface is such that in-plane emission 
is prefered. It is therefore also the interplay between the 
timescales of passing time of the spectators and expansion time of 
the dense, stopped interaction zone which determines 
the time-integrated elliptic flow signal. 
Indeed, when following the elliptic flow as a function of 
reaction time, early out-of-plane squeeze is superposed by later 
preferential in-plane expansion \cite{sorge97}. 
So, the sign of the elliptic flow changes twice as a function of 
incident energy: At intermediate 
energies ($E_{\rm lab} \approx 100 \,A\,$MeV) 
a change from in-plane emission (rotation-like behaviour) to the squeeze-out is 
predicted \cite{soff95,crochet97} where at relativistic energies (
$E_{\rm lab} \approx 5 \,A\,$GeV) the opposite change from 
the squeeze-out to in-plane enhancement is observed.

\begin{figure}[htp]
\vspace*{-.5cm}
\centerline{\hspace{.8cm}\hbox{
\epsfig{figure=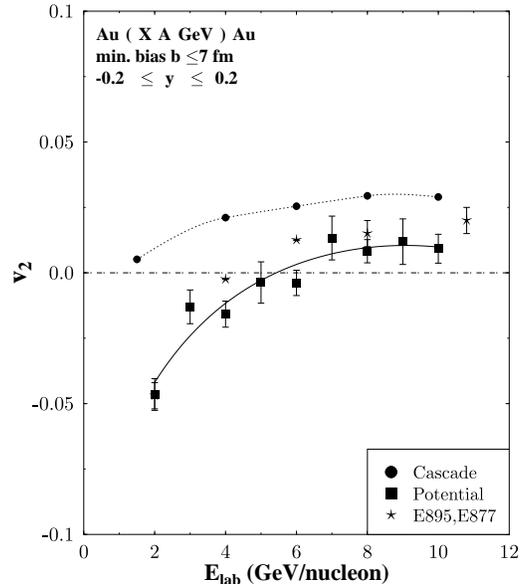,width=7cm}}}
\caption{Elliptical flow parameter $v_2$ as a function of incident
energy $E_{\rm lab}$ for midrapidity protons in Au+Au collisions. 
Data of the E895 and E877 collaborations (stars) and UrQMD calculations with (squares)
and without (circles) mean field potentials are displayed.}
\end{figure}

Fig.~1 shows the excitation function of the in-plane/squeeze-out 
flow parameter $v_2$. This is observed by $90^{o}$ peaks in the 
azimuthal angular distribution 
$dN/d\Phi$ \cite{stoecker82a,gut89,crochet97,leifels93a,brill96a} 
of nucleons at midrapidity 
for Au+Au collisions with the Fourier expansion 
\begin{equation}
\frac{dN}{d\Phi}=v_0\left(1+ 2\,v_1 \,\cos (\Phi) + 2\,v_2 \, \cos(2\Phi)\right)
\,. 
\end{equation}
$v_0$ is for normalization only, where $v_1$ characterizes the 
directed in-plane flow. 
While $v_2 > 0$ indicates in-plane enhancement, $v_2 < 0$ characterizes the 
squeeze-out perpendicular to the event plane.
Data by the E895 \cite{pinkenburg99} and the E877 \cite{PBM98} collaborations (stars) and 
UrQMD calculations are displayed. The UrQMD calculations are performed 
within the cascade mode (circles) as well as with mean field 
potentials (squares).  
A detailed survey on the UrQMD model and its underlying concepts is available \cite{bass98}. 
Clearly, the experimental observation of 
a transition from squeeze-out to a preferential in-plane emission can 
only be described with the potentials included. 
The cascade simulations do not show the squeeze-out due to 
the lack of the strongly repulsive nucleonic potential at this energy. 
The data are consistently described with potentials corresponding 
to an equation of state with {\it one} incompressibility ($K=380\,{\rm MeV}$), independent 
on the incident energy. This is in contrast to findings in 
\cite{pinkenburg99} where a softening of the equation of state with incident 
energy is deduced from the comparison to transport model calculations \cite{danielewicz98a}. 

Transverse flow has been discovered even at the highest energies 
at the SPS for the Pb+Pb system at 158$\,A\,$GeV both by the 
NA49 and by the WA98 collaborations. 
Here, UrQMD calculations are compared to the flow parameters 
$v_1$ and $v_2$, which can also be expressed by \cite{appelshaeuser98}
\begin{equation}
v_1 = \langle \frac{p_x}{p_t} \rangle ,  
\hspace{1cm} v_2=\langle \left( \frac{p_x}{p_t} \right) ^2 -
\left( \frac{p_y}{p_t} \right) ^2\rangle\, .
\end{equation}

Fig.2 shows the rapidity dependence of the proton flow (upper half) 
and of the flow of charged pions (lower half). 
Full symbols are UrQMD calculations  where open symbols are experimental 
data \cite{appelshaeuser98}. 
The data are reflected at midrapidity ($y_{\rm lab} \approx 2.9$). 
In reflection the signs of the $v_1$ values have been reversed 
in the backward hemisphere, but not the $v_2$ values \cite{appelshaeuser98}.
For the directed transverse flow ($v_1$), both data as well as UrQMD results 
exhibit  a characteristic S-shaped curve.  
The elliptic flow values ($v_2$) seem to be slightly peaked at 
medium rapidity ($y_{\rm lab}\approx 4-4.5$ and 
$y_{\rm lab}\approx 1.5-2$), both for pions and protons, contrary
to what was inferred in \cite{appelshaeuser98}. 
Both protons and pions show an in-plane enhanced emission ($v_2 > 0$). 
\begin{figure}[htp]
\vspace*{-.5cm}
\centerline{\hspace{.8cm}\hbox{\epsfig{figure=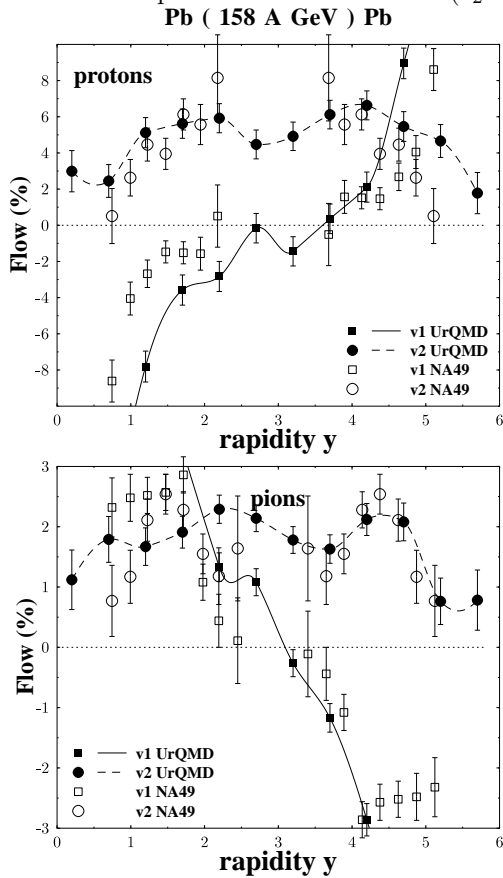,width=7cm}}}
\caption{Flow parameters $v_1$ and $v_2$ as a function of rapidity for protons
(upper diagram) and charged pions (lower diagram). Open symbols are data and full symbols 
display the UrQMD results}
\label{slopes}
\end{figure}
The proton flow shows positiv flow whereas the pion flow exhibits 
the opposite negative sign, caused by absorption and rescattering effects.  
The overall agreement between data and calculations looks rather good. 
Discrepancies are seen dominantly for the high rapidity pion 
directed flow ($v_1$), which is too strong in the calculations compared 
to the data which show saturation of $v_1$ for $y_{\rm lab} > 4$ and 
$y_{\rm lab} < 2$. Also the proton directed flow seems to be 
slightly too strong at high rapidity. 
The elliptic flow shows good agreement for the sign as well as for the 
magnitude of $v_2$ ($v_2\approx 5 \%$ for protons and $v_2 \approx 2\% $
 for pions).

\begin{figure}[htp]
\vspace*{-.5cm}
\centerline{\hspace{.8cm}\hbox{\epsfig{figure=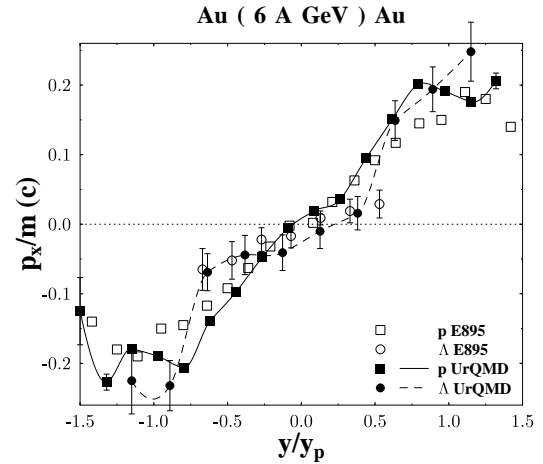,width=7cm}}}
\caption{Directed transverse flow $p_x/m$ as a function of normalized rapidity $y/y_p$ for 
protons (squares) and $\Lambda$'s (circles). Open symbols are data and full symbols are 
the results of UrQMD calculations.}
\end{figure}

Strong directed flow has also been discovered in the energy region where the elliptic 
flow disappears. Fig.~3 shows the directed transverse flow $p_x/m$ as a function  
of the normalized rapidity for protons (squares) and $\Lambda$'s (circles) in 
Au$(\,6\,A\,{\rm GeV}\,)$Au collisions. 
Open symbols are preliminary data by the E895 collaboration \cite{pinkenburg99} 
and full symbols display the results of UrQMD calculations. 
The proton data are reflected at midrapidity. 
Both, protons and $\Lambda$'s show strong positive directed flow. 
The proton flow is larger than the $\Lambda$ flow close to midrapidity 
($|y/y_p| \le 0.6$) both in the data as well as in the UrQMD calculations. 
At target/projectile rapidity the $\Lambda$ flow is predicted to exhibit  
a similar magnitude as the protons show. 
The species-dependent flow pattern clearly demonstrates a 
complex non-hydrodynamic behaviour which seems to rule out simple 
fireball+flow models.

In summary, recent data on the collective flow in heavy ion collisions 
at the SPS and AGS have been compared to UrQMD calculations. 
The excitation function of the elliptic flow at midrapidity for the Au+Au system 
shows a transition from 
the squeeze-out to an in-plane enhancement. The data agree with the 
calculations done with an equation of state with {\it one} incompressibility. 
Therefore a softening of the equation of state cannot be deduced 
from this comparison. 

The elliptic flow at the SPS for Pb+Pb collisions shows in-plane enhancement, both for 
protons and pions in the full rapidity range. The UrQMD results show complete 
agreement to data.
The positive directed flow of protons is opposite to the directed flow of pions 
which show an anti-flow. While good agreement exists around midrapidity,  
the pion flow is too strong in the calculations at high rapidities. 
This seems to be due to the high momentum tails of the pion 
transverse momentum distribution and will be investigated in a forthcoming publication. 
The directed proton flow also seems to be slightly overestimated at high rapidities 
by the UrQMD results.
 
The comparison of Au$(\,6\,A\,{\rm GeV}\,)$Au collisions demonstrates that strong 
directed flow is present for protons and $\Lambda$'s, where the $\Lambda$'s 
show less flow than the protons around midrapidity. 
At higher rapidities the $\Lambda$ flow is predicted to show similar magnitude 
as the proton flow. 

The species-dependent flow patterns 
illustrate the complex collision dynamcis and demonstrate the necessity
of highly non-trivial microscopic transport models for an adequate 
description of relativistic heavy ion collisions.

\acknowledgements
This work has been
supported in part by BMBF, DFG, GSI and Graduiertenkolleg 
'Experimentelle und Theoretische Schwerionenphysik'.
S.A.B.\ is supported in part by the Alexander von Humboldt Foundation
through a Feodor Lynen Fellowship, and by DOE grant DE-FG02-96ER40945.
S.~S. and M.~B. thank the Josef Buchmann Foundation for support.


\begin{references}

\bibitem{pinkenburg99}
C.~Pinkenburg et al. (E895 Collaboration), 
\newblock e-Print Archive: nucl-ex/9903010, and 
P.~Chung et al. (E895 Collaboration),
J.~Phys.~G: Nucl.~Part.~Phys.~{\bf 25}, 255 (1999).

\bibitem{appelshaeuser98}
H.~Appelsh\"auser et al. (NA49 Collab.), 
Phys.~Rev.~Lett.~{\bf 80}, 4136 (1998), and 
A.~M.~Poskanzer and S.~A.~Voloshin,
Phys.~Rev.~{\bf C58}, 1671 (1998), and 
S.~A.~Voloshin,
Phys.~Rev.~{\bf C55}, R1630 (1997).

\bibitem{aggarwal98}
M.~M.~Aggarwal et al. (WA98 Collaboration),
\newblock e-Print Archive: nucl-ex/9807004 v2.

\bibitem{scheid68a}
\mbox{W.~Scheid, R.~Ligensa and W.~Greiner,} 
Phys.~Rev.~Lett.~{\bf 21}, 1479 (1968) and 
Phys.~Rev.~Lett. {\bf 32}, 741 (1974).

\bibitem{hofmann76a}
        J.~Hofmann et al., 
        Phys.~Rev.~Lett.~{\bf 36}, 88 (1976).

\bibitem{stoecker86a}
        H.~St\"ocker and W.~Greiner, 
        Phys.~Rep.~{\bf 137}, 277 (1986).

\bibitem{stoecker81a}
        H.~St\"ocker and B.~M\"uller, 
        LBL-preprint~12471, unpublished.

\bibitem{stoecker80a}
        H.~St\"ocker, J.~A.~Maruhn and W.~Greiner, 
        Phys.~Rev.~Lett.~{\bf 44}, 725 (1980).

\bibitem{stoecker82a}
        H.~St\"ocker et al.,
        Phys.~Rev~{\bf C25}, 1873 (1982).

\bibitem{buchwald84a}
        G.~Buchwald et al., 
        Phys.~Rev.~Lett.~{\bf 52}, 1594 (1984).

\bibitem{schmidt93a}
        W.~Schmidt et al., 
        Phys.~Rev.~{\bf C47}, 2782 (1993).

\bibitem{moli84}
        J.~J.~Molitoris et al., 
        Phys.~Rev.~Lett.~{\bf 53}, 899 (1984).

\bibitem{moli85}
        J.~J. Molitoris and H.~St\"ocker, 
        Phys.~Lett.~{\bf B162}, 47 (1985).

\bibitem{gyulassy82}
        M.~Gyulassy, K.~A.~Frankel, H.~St\"ocker, 
        Phys.~Lett.~{\bf B110}, 185 (1982).

\bibitem{hartnack89a}
        C.~Hartnack et al.,
        Nucl.~Phys.~{\bf A495}, 303c (1989).

\bibitem{ch92}
        C.~Hartnack et al.,
        Nucl.~Phys.~{\bf A538}, 53c (1992) and 
        Mod.~Phys.~Lett.~{\bf A9}, 1151 (1994).

\bibitem{peilert94}  
         G.~Peilert, H.~St\"ocker, W,~Greiner, 
       \newblock Rep.~Prog.~Phys.~{\bf 57}, 533 (1994).

\bibitem{gus84}
        H.-A.~Gustafsson et al.,
        Phys.~Rev.~Lett.~{\bf 52}, 1590 (1984).

\bibitem{do86}
        K.~G.~R. Doss et al.,
        Phys.~Rev.~Lett.~{\bf 57}, 302 (1986).

\bibitem{gut89}
H.~H.~Gutbrod et al.,
\newblock  Phys.~Lett.~{\bf B216}, 267 (1989)  
and Phys.~Rev.~{\bf C42}, 640 (1990).

\bibitem{renfordt84a}
        R.~Renfordt~et~al.,
        Phys.~Rev.~Lett.~{\bf 53}, 763 (1984).

\bibitem{gosset90a}
        J.~Gosset~et~al.,
        Phys.~Lett.~{\bf B247}, 233 (1990).

\bibitem{dem90}
M.~Demoulins et al.,
\newblock  Phys.~Lett.~{\bf B241}, 476 (1990)
and Phys.~Rev.~Lett.{\bf ~62}, 1251 (1989).

\bibitem{ramillien95a}
        V. Ramillien et~al.,
        Nucl.Phys.~{\bf A587}, 802 (1995).

\bibitem{herrmann96a}
        N.~Herrmann~et~al.,
        Nucl.~Phys.~{\bf A610}, 49 (1996).

\bibitem{leifels93a}
        Y.~Leifels~et~al.,
        Phys.~Rev.~Lett.~{\bf 71}, 963 (1993).

\bibitem{kugler94a}
        A.~Kugler~et~al.,
        Phys.~Lett.~{\bf B335}, 319 (1994).

\bibitem{brill96a}
        D.~Brill~et~al.,
        Z.~Phys.~{\bf A357}, 207 (1997).

\bibitem{chance96a}
        J.~Chance~et~al.,
        Phys.~Rev.~Lett.~{\bf 78}, 2535 (1997).
        
\bibitem{westfall93}
        G.~D.~Westfall et al.,
        Phys.~Rev.~Lett.~{\bf 71}, 1986 (1993).

\bibitem{ollitrault93a}
        J.~Y.~Ollitrault,
        Phys.~Rev.~{\bf D48}, 1132 (1993).

\bibitem{barrette94c}
        J.~Barrette~et~al.,
        Phys.~Rev.~Lett.~{\bf 73}, 2532 (1994).

\bibitem{barrette96a}
        J.~Barrette~et~al.,
        Phys.~Rev.~{\bf C55}, 1420 (1997).

\bibitem{liu98a}
H.~Liu et al. (E895 Collaboration),
\newblock Nucl.~Phys.~{\bf A638}, 451 (1998).

\bibitem{ogilvie98a}
        C.~Ogilvie~et~al.,
        Nucl.~Phys.~{\bf A638}, 57 (1998).

\bibitem{keitz91}
A.~v.~Keitz et al.,
\newblock Phys.~Lett.~{\bf B263}, 353 (1991).

\bibitem{nishimura98a}
        S.~Nishimura~et~al.,
        Nucl.~Phys.~{\bf A638}, 549 (1998).

\bibitem{vanhove82a}
        L.~van~Hove,
        Phys.~Lett.~{\bf B118}, 138 (1982).


\bibitem{amelin91}
        N.~S.~Amelin, et al., 
        Nucl~Phys.~{\bf A544}, 463c (1992), and L.~V.~Bravina et al., 
        Nucl.~Phys.~{\bf A566}, 461c (1994) and 
        Phys.~Lett.~{\bf B344}, 49 (1995).

\bibitem{rischke95b}
        D.~H.~Rischke, Y.~P\"urs\"un, J.~A.~Maruhn, H.~St\"ocker and W.~Greiner,
        Heavy~Ion~Physics~{\bf 1}, 309 (1995).

\bibitem{brachmann97}
        J.~Brachmann, et al.,
        Nucl.~Phys.~{\bf A619}, 391 (1997).

\bibitem{shuryak95}
C.~M.~Hung, E.~V.~Shuryak, 
\newblock  Phys.~Rev.~Lett.~{\bf 75}, 4003 (1995),

\bibitem{kapusta85}
        J.~Kapusta, S.~Pratt, L.~Mc~Lerran and H.~v.~Gersdorff, 
        Phys.~Lett.~{\bf B163}, 253 (1985).

\bibitem{gersdorf87}
        H.~v.~Gersdorff, 
        Nucl.~Phys.~{\bf A461}, 251c (1987).

\bibitem{rischke96a}
        D.~H.~Rischke and M.~Gyulassy, 
\newblock Nucl.~Phys.~{\bf A597}, 701 (1996).

\bibitem{sorge97}
        H.~Sorge,
\newblock Phys.~Rev.~Lett.~{\bf 78}, 2309 (1997).
        
\bibitem{soff95}
        S.~Soff et al.,
Phys.~Rev.~{\bf C51}, 3320 (1995).

\bibitem{crochet97}
P.~Crochet et al. (FOPI Collab.),
\newblock Nucl.~Phys.~{\bf A624}, 755 (1997).

\bibitem{PBM98}
P.~Braun-Munzinger and J.~Stachel, 
   Nucl.~Phys.~{\bf A 638}, 3c (1998).

\bibitem{bass98}
S.~A.~Bass et al., 
Prog.~Part.~Nucl.~Phys.~{\bf 41}, 225 (1998).

\bibitem{danielewicz98a} 
        P.~Danielewicz et al., 
        Phys.~Rev.~Lett.~{\bf 81}, 2438 (1998).
        
\end{references}
\end{document}